\documentclass[aps,twocolumn,showpacs,amssymb]{revtex4}
\usepackage{graphicx}
\begin{document}

\title{How periodic orbit bifurcations drive multiphoton ionization}

\author{S. Huang$^1$}
\email{gtg098n@mail.gatech.edu}
\author{C. Chandre$^2$}
\author{T. Uzer$^1$}

\affiliation{$^1$ Center for Nonlinear Science, School of Physics,
Georgia Institute of Technology, Atlanta, Georgia 30332-0430, U.S.A.\\
$^2$ Centre de Physique Th\'eorique - CNRS, Luminy - Case 907, 13288 Marseille
cedex 09, France}

%\date{\today}

\begin{abstract}
The multiphoton ionization of hydrogen by a strong bichromatic microwave field is a complex process prototypical for atomic control research. Periodic orbit analysis captures this complexity: Through the stability of periodic orbits we can match qualitatively the variation of experimental ionization rates with a control parameter, the relative phase between the two modes of the field. Moreover, an empirical formula reproduces quantum simulations to a high degree of accuracy. This quantitative agreement shows how short periodic orbits organize the dynamics in multiphoton ionization.
\end{abstract}


\maketitle

Simple systems can display extraordinarily complex dynamics: This lesson from three decades of chaos theory has altered the direction of many areas of physics \cite{chaosbook}. One-electron  systems, which are the most fundamental at the atomic level, have had their fair share of such striking new discoveries \cite{blumel9}, a prominent example being the  multiphoton ionization of hydrogen in a strong microwave field \cite{Bayfield}. Its interpretation remained a puzzle until its stochastic, diffusional nature was uncovered through the then-new theory of chaos \cite{Meerson}. Feverish research activity in the last thirty or so years has resulted in a fairly complete understanding of this problem \cite{Casati5,Jenson,kochreview}. As in many other areas of physics, here too the emphasis has shifted lately from understanding the process to using those insights to manipulating it \cite{rabitz,shapiro}. Clearly, the hope is that experience gained from this prototypical system will help to control more complex systems ranging from atoms to plasmas. Here control denotes tailoring the physical behavior of nonlinear dynamical systems (which generically exhibit chaotic dynamics) using ``knobs'' (i.e., suitable external parameters). Success depends on identifying simple knobs and understanding why and how they alter the system. In this Letter we report on precisely such a knob for a complex quantum system, and show how ionization behavior can be predicted with quantitative accuracy using periodic orbits.

The ionization of a one-dimensional hydrogen atom driven by a bichromatic linearly polarized electric field is a seemingly simple set-up with complex dynamics. Bichromatic pulses \cite{ko,Ehlotzky} have emerged as natural tools in atomic control research because the relative phase of the two phase-locked pulses offers a practical control knob \cite{Howard,Haffmans,ivanov,Buchleitner,petrosyan,Sirko1,Sirko,batista,PMKoch,rangan,greeks}. The Hamiltonian of the system is, in atomic units,
\begin{equation}
\label{Hatom} H=\frac{p^{2}}{2}-\frac{1}{x}+F_{h} x\sin(h\omega
t)+F_{l} x\sin (l\omega t+\phi),
\end{equation}
where the indices $l, h$ refer to the low and high frequency modes with frequencies $l\omega$ and $h\omega$, respectively.
These two modes are frequency locked, i.e.\ $l$ and $h$ are
integers. They are out-of-phase by $\phi$, the control parameter. 

In what follows, we consider Hamiltonian~(\ref{Hatom}) for the two sets of experimental parameters of Ref.~\cite{PMKoch}: $(I)$ the $h$:$l$=3:1, $F_h=24\mbox{ Vcm}^{-1}$ and $F_l=53.4\mbox{ Vcm}^{-1}$, and $(II)$ the $h$:$l$=3:2, $F_h=25\mbox{ Vcm}^{-1}$ and $F_l=33.5\mbox{ Vcm}^{-1}$. The highest frequency is $18\mbox{ GHz}$ in both cases. These
two sets show experimentally and numerically drastically different
ionization behavior~\cite{PMKoch} and in the frequency range considered,
these results can be reproduced by quantum or classical
simulations.
Our purpose here is to show that these findings can be qualitatively and quantitatively captured using a periodic orbit analysis, which reveals the classical bifurcations responsible for ionization. It also allows prediction of ionization at other values of parameters without resorting to large numerical simulations.

We begin by mapping Hamiltonian~(\ref{Hatom}) into
action-angle variables
such that the principal quantum number $n$ is associated with action
$J$. We assume $\omega=1$ without loss
of generality. Hamiltonian~(\ref{Hatom}) becomes~\cite{Casati5}
\begin{eqnarray}
H&=&-\frac{1}{2J^{2}}+2J^{2}[F_{h}\sin(ht)\nonumber
\\ &&+F_{l}\sin (l t+\phi)]\left( a_{0}/2+\sum_{k=1}^\infty{a_{k}\cos
k\theta}\right),\label{HatomAA}
\end{eqnarray}
where $a_{n}=[J_{n}(n)-J_{n-1}(n)]/n$ and $J_{n}$'s are Bessel
functions of the first kind. Note that there are three variables
$(J,\theta,t)$ and three parameters $(F_h,F_l,\phi)$. We denote this Hamiltonian $H(J,\theta,t;\phi)$ where $(F_h,F_l)$ are chosen according to Cases $(I)$ or $(II)$.

For this Hamiltonian, we consider a specific periodic orbit, denoted ${\mathcal O}(0)$ for $\phi=0$. Numerically it is
determined using a modified Newton-Raphson multi-shooting
algorithm as described in Ref.~\cite{chaosbook}. We follow
this periodic orbit as $\phi$ is varied. The orbit ${\mathcal
O}(0)$ deforms continuously into ${\mathcal O}(\phi)$, the period
of which is denoted $T(\phi)$. In addition to its location, we also monitor
its linear stability properties obtained from the reduced tangent
flow expressed as
$$
\frac{d{\mathcal J}_\phi^t}{dt}={\mathbb J}\nabla^2
H(J,\theta,t;\phi) {\mathcal J}^t_\phi,
$$
where ${\mathbb J}$ is the two-dimensional skew-symmetric matrix
and $\nabla^2 H$ is the two-dimensional Hessian matrix (composed by second
derivatives of $H$ with respect to its canonical variables $J$ and
$\theta$). The initial condition is ${\mathcal J}_\phi^0={\mathbb
I}_2$ (the two-dimensional identity matrix). The stability
properties are given by the two eigenvalues of the monodromy
matrix ${\mathcal J}_\phi^{T(\phi)}$ which form a pair
$(\lambda,1/\lambda)$ (since the flow is volume preserving, the
determinant of ${\mathcal J}_\phi^{T(\phi)}$ is equal to 1). The
periodic orbit is elliptic if the spectrum is $({\mathrm
e}^{i\omega(\phi)},{\mathrm e}^{-i\omega(\phi)})$ (and stable,
except at some particular values), or hyperbolic if the spectrum
is $(\lambda(\phi),1/\lambda(\phi))$ with
$\lambda(\phi)\in{\mathbb R}^*$ (unstable). In a more concise
form, it can be summarized using Greene's
residue, $R$~\cite{gree79,mack92}
$$
R(\phi)={(2-\mbox{tr}{\mathcal J}_\phi^{T(\phi)})}/{4}.
$$
If $R(\phi)\in ]0,1[$, the periodic orbit is elliptic; if
$R(\phi)<0$ or $R(\phi)>1$ it is hyperbolic; and if $R(\phi)=0$
and $R(\phi)=1$, it is parabolic. Generically, periodic orbits and
their linear stabilities are robust against small changes of parameters,
except at specific values where bifurcations occur \cite{Cary}. These
rare events affect the dynamical behavior drastically.

In what follows, we identify the bifurcations (if any) of a set of short periodic orbits, i.e., the
type and the value of the parameter $\phi_c$ where they bifurcate.
This provides a way to foretell if a relatively high ionization rate
should be expected or not. We use the residue curves $\phi \mapsto R(\phi)$ for each
periodic orbit to analyze the dependence of ionization rates on $\phi$ .  The importance of considering two associated Birkhoff periodic
orbits (i.e.\ periodic orbits with the same action but different
angles in the integrable case, one elliptic and one hyperbolic), was emphasized in Ref.~\cite{Bachelard}

\begin{figure}
 \centering
 \includegraphics[width=8.5cm,height=7.5cm]{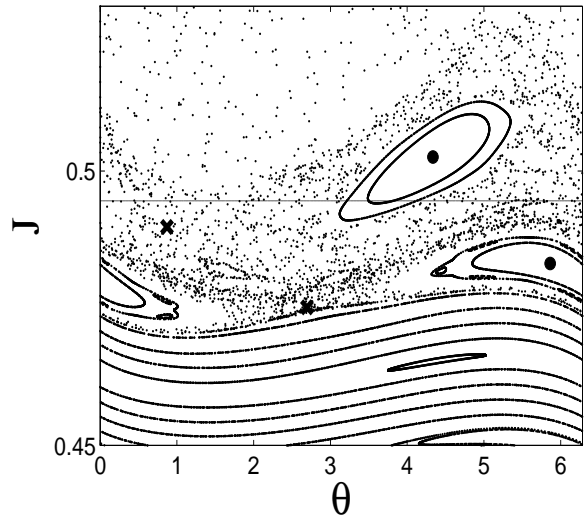}
\caption{\label{fig:fig1}Poincar\'e section of
Hamiltonian~(\ref{HatomAA}) for Case $(I)$ at $\phi=0$. Full
circles and crosses indicate the two elliptic and two hyperbolic periodic orbits we consider respectively. The
horizontal line corresponds to the principal quantum number $n=51$
considered in Ref.~\cite{PMKoch}.}
\end{figure}

\begin{figure}
 \centering
 \includegraphics[width=8.5cm,height=7.5cm]{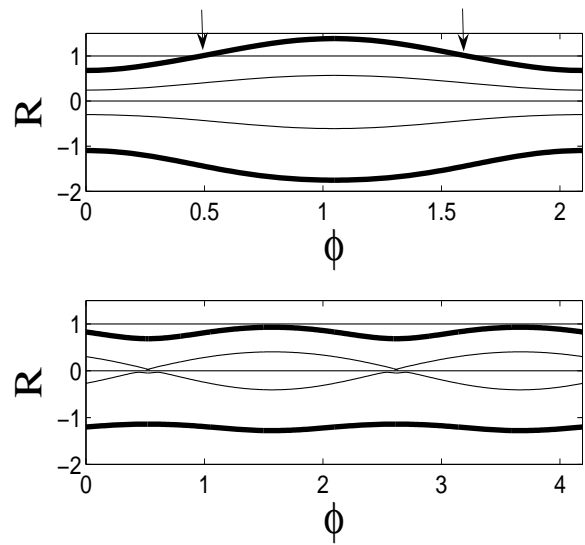}
\caption {\label{fig:fig2} Residue curves for the four periodic
orbits with period $2\pi$ indicated by crosses and circles on
Fig.~\ref{fig:fig1}. The bold curves are for the upper set of
elliptic/hyperbolic orbits. Small arrows indicate where the
bifurcations happen. Upper panel is for Case $(I)$, i.e.\ $h$:$l$=3:1 and lower panel for Case $(II)$, i.e.\ $h$:$l$=3:2.}
\end{figure}

Figure~\ref{fig:fig1} shows a Poincar\'e section of
Hamiltonian~(\ref{HatomAA}) for Case $(I)$ at $\phi=0$. We notice two main islands in the chaotic sea. At the
centers of these islands sit elliptic periodic orbits with
period $2\pi$ (indicated by full circles). Note that the
(rescaled) principal quantum number considered in
Ref.~\cite{PMKoch} lies in between these two islands. The residue
method will monitor these two elliptic periodic orbits as well as
their associated hyperbolic periodic orbits (they all have the same period
$T(\phi)=2\pi$ for all values of $\phi$)~\cite{Bachelard}. We
follow the location and residue for each of these periodic orbits as the parameter $\phi$ varies.
Figure~\ref{fig:fig2} shows the residue curves of these orbits.

In Fig.~\ref{fig:fig2} (upper panel), if we start from $\phi=0$ and follow the
upper elliptic periodic orbit of Fig.~\ref{fig:fig1}, it remains
elliptic ($R(\phi)\in ]0,1[$) until $\phi_c\approx 0.49$ where a
bifurcation appears. At this critical value, the orbit is
parabolic. Then it turns hyperbolic ($R(\phi)> 1$) until
$2\pi/3-\phi_c$ where another bifurcation occurs and the
orbit returns to being elliptic. This bifurcation is of the period
doubling kind at $\phi_c$ and a period halving one at $2\pi/3-\phi_c$. The computation of $\log|\lambda_\pm(\phi)|$ (where
$\lambda_\pm(\phi)$ are the two eigenvalues of the monodromy
matrix ${\mathcal J}_\phi^{T(\phi)}$ associated with the upper
elliptic periodic orbit) show that, before the bifurcation, $\log|\lambda(\phi)|=0$
since the orbit is elliptic, and that just after the bifurcation, $\log|\lambda_\pm(\phi)|\propto \pm \sqrt{\phi-\phi_c}$.
Note that while the upper
elliptic periodic orbit undergoes a bifurcation as $\phi$ is
varied, the other three periodic orbits retain the stability
properties they had at $\phi=0$. 

In the parameter range $\phi\in ]\phi_c,2\pi/3-\phi_c[$, the upper part of phase
space does exhibit more chaos. Since the initial atomic beam is taken in this region (principal
quantum number $n=51$ which corresponds to action $J_i\approx
0.495$), the ionization rate is expected to be
higher in this regime. Apart from fine detail (like higher
order islands), the upper part of phase space is roughly
homogeneous. Therefore we expect to observe a plateau in the
ionization probability versus $\phi$. The reason is that a strongly hyperbolic orbit only influences the ionization time and not the value of the ionization probability. Of course, this is true provided that the duration of the maximum pulse envelope is large enough (in the experiment, this is approximately 15 times the period of the periodic orbits considered~\cite{PMKoch}). Roughly speaking, this means that in the chaotic region all the orbits ionize (i.e.,\ escape to a value of the action $J_\mathrm{ion}\gtrsim 1.26$) whatever the hyperbolicity degree is. 
In Ref.~\cite{PMKoch},
experimental results as well as one-dimensional quantum
calculations show this plateau. From quantum calculations, $\phi_c\approx 0.5$ was obtained in Ref.~\cite{PMKoch} which is in good agreement with the parameter value $\phi_c\approx 0.49$ at which the bifurcation of the upper elliptic periodic orbit
occurs.

We perform the same analysis for Case $(II)$ where $h$:$l$=3:2.
Again we consider two main islands in the chaotic sea where two different
elliptic periodic orbits with period $2\pi$ sit at the centers. We
monitor the stability of four periodic orbits (two elliptic
orbits and the associated hyperbolic ones).
Figure~\ref{fig:fig2} (lower panel) shows the four residue curves. Evidently
 the elliptic periodic orbits remain
elliptic and the hyperbolic ones remain hyperbolic for all values of $\phi$. No
bifurcations occur; consequently, the ionization probability is expected to be approximately independent of $\phi$ and to be lower than Case
$(I)$ since for these values of amplitudes, the chaotic
region is smaller. This is consistent with the experimental and quantum
calculations of Ref.~\cite{PMKoch} (see their Fig.~7).
The experimental results show a nearly flat curve for the
ionization probability versus $\phi$, whereas the
quantum calculations show significant variations for this
probability but no sharp increase and decrease as in Case $(I)$.

The periodic orbit analysis above elucidates whether or not
there is a significant ionization probability for specific
parameter values, and also where plateaus are
expected to occur. This qualitative agreement highlights the important role played by these orbits. Furthermore, we can obtain quantitative agreement concerning the shape of
the ionization curve versus the phase $\phi$ by using the residue
curves. To this end, we devise an empirical formula for relative
ionization probability in the
following way~: First, the values of $\phi$ giving the highest
ionization would be the ones associated with the highest variations of
the residues (in absolute value) with respect to the minimum ionization. 
Second, if the periodic orbit is too
far (in action) from the considered action $J_i$ then it will not
influence the dynamics so there should be a penalizing
term depending on its position with respect to the chosen rescaled
action. The formula reads~:
\begin{equation}
\label{Ionempirical}
P_{\mathrm{ion}}(\phi)=A+B\sum_{m=1}^{M}\frac{\exp |R_{m}(\phi)-R_{m}(0)|}{\exp |\overline{J_{m}(\phi)}-J_{i}|},
\end{equation}
where the sum is taken over the $M$ different periodic orbits considered and $\overline{J_{m}(\phi)}=\int_{0}^{2\pi}J(\theta)d\theta/2\pi$ is the average action of the periodic orbit $m$. The parameters $A$ and $B$ in Eq.~(\ref{Ionempirical}) are merely a translation and a dilatation of the curve in order to match the mean value and the amplitude of variations of $P_{\mathrm{ion}}$ obtained in Ref.~\cite{PMKoch}. This formula takes into account the value of the residues at $\phi=0$. The aim is to set up a baseline for each of the periodic orbits (which is taken here at the value of the parameter where the ionization is minimal~\cite{PMKoch}). In general, Eq.~(\ref{Ionempirical}) can exhibit values which are greater than 1, which are not relevant. In order to remedy to this problem, we truncate $P_{\mathrm{ion}}$ at the value where a bifurcation occurs. Therefore in the range where $R_n(\phi)$ is larger than one, $P_{\mathrm{ion}}$ is constant (taken as the value of the residue at $\phi_c$ where the bifurcation occurs). 

\begin{figure}
 \centering
 \includegraphics[width=8.5cm,height=7.5cm]{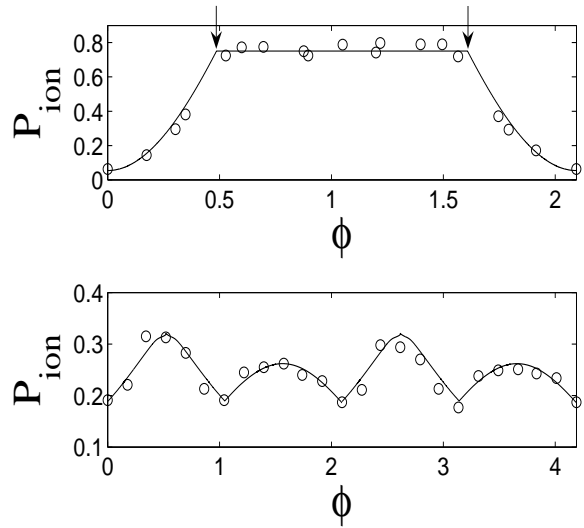}
 \caption{\label{fig:fig3} Normalized ionization rate {\em vs} $\phi$ based on
Eq.~(\ref{Ionempirical}) for Case $(I)$ (with $h$:$l$=3:1) with
$A=-2.52$ and $B=0.65$ (upper panel), and for Case $(II)$ (with $h$:$l$=3:2) with $A=-0.485$ and $B=0.17$. Circles represent the data obtained by one-dimensional quantum calculations, taken from Ref.~\cite{PMKoch}.}
\end{figure}

Figure~\ref{fig:fig3} depicts $P_{\mathrm{ion}}$ given by Eq.~(\ref{Ionempirical}) versus parameter $\phi$ as well as the data taken from Ref.~\cite{PMKoch} for both cases [Case $(I)$ (upper panel) and Case $(II)$ (lower panel)]. Since there are no bifurcations in Case $(II)$, there is no plateau. We notice that the empirical formula reproduces accurately the results obtained from quantum calculations. In particular, it captures some essential features of the ionization curve in Case $(II)$, like the two unequal-sized peaks and the specific shape of both peaks. 

\begin{figure}
  \centering
  \includegraphics[width=8.5cm,height=7.5cm]{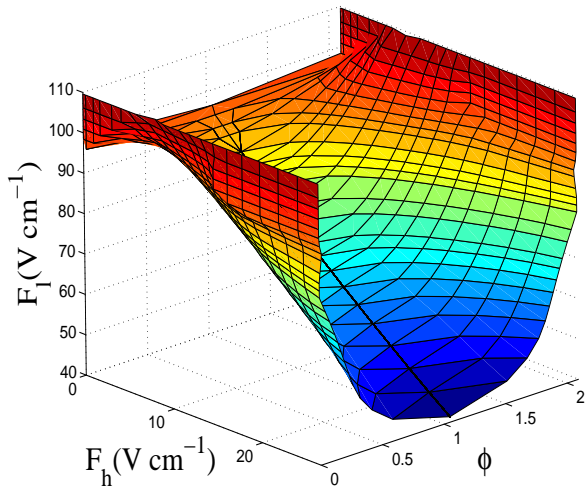}
  \caption{\label{fig:fig4} Bifurcation surface in parameter space $(\phi, F_h, F_l)$ for $h$:$l$=3:1.}
\end{figure}

We can also predict the behavior of the system as all three parameters (the two amplitudes and the relative phase of the field) are varied. In Fig.~\ref{fig:fig4}, we represent the set of parameters where the upper elliptic periodic orbit of Fig.~\ref{fig:fig1} (in the 3:1 case) is in fact parabolic (i.e.,\ the set of parameters where the system undergoes a major bifurcation). The equation of this surface in parameter space is $R(\phi, F_h, F_l)=1$. The boundaries of the plateaus in parameter $\phi$ of Fig.~\ref{fig:fig3} (upper panel) obtained by fixing the two values for $F_h$ and $F_l$ are on this surface. We notice that as expected, when $F_{h}$ approaches zero, this surface is less dependent on parameter $\phi$. Table~\ref{tab1} compares the experimental values for ionization thresholds~\cite{Sirko} with corresponding values from our bifurcation analysis. 

\begin{table}
\caption{\label{tab1}Ionization thresholds obtained for $F_h=6\mbox{ Vcm}^{-1}$, experimentally in Ref.~\cite{Sirko}  and by the residue method (see Fig.~\ref{fig:fig4}). The 1f case corresponds to $F_h=0$.}
\begin{ruledtabular}
\begin{tabular}{c|ccc}
$F_l (\mbox{ Vcm}^{-1})$ & $\phi=0$ & $\phi=\pi/3$ & 1f\\
\hline
\cite{Sirko} & 107 & 85 & 96\\
residue & 109.6 & 81.4 & 96.2
\end{tabular}
\end{ruledtabular}
\end{table}

To obtain this ionization surface in parameter space, a large number of classical trajectories for each value of the parameters $(\phi,F_h,F_l)$ needs to be computed for a sufficiently long time in order to decide whether or not a given trajectory has ionized. In contrast, only one orbit for a short time (typically the period of the field) is needed for the residue analysis. Furthermore, using residues, this surface can be constructed locally without any need to consider all possible values of the parameters. 

More broadly, our results constitute rules by which this quantum system can be controlled. Such systematic and practical coherent control rules remain very rare and sought after \cite{rabitz}.

This research was supported by the US National Science Foundation.
C.C. acknowledges support from Euratom-CEA (contract EUR~344-88-1~FUA~F).

\end{document}